\documentstyle[aps,prb,multicol,epsf]{revtex}
\newcommand{\startlongequation}{
\end{multicols}\vspace*{-3.5ex}{\tiny\noindent
\begin{tabular}[t]{c|} \parbox{0.493\hsize}{~} \\ \hline \end{tabular}} }
 
\newcommand{\stoplongequation}{
{\tiny\hspace*{\fill}
\begin{tabular}[t]{|c}\hline\parbox{0.49\hsize}{~} \\ \end{tabular}}
\vspace*{-2.5ex}\begin{multicols}{2} }
 
\begin{document}
\draft
\title{Anomalous relaxations and chemical trends
at  III-V nitride non-polar  surfaces}
\author{Alessio Filippetti,$^{1,2}$  Vincenzo Fiorentini,$^{1,3}$
 Giancarlo Cappellini,$^1$ and Andrea Bosin$^4$}
\address{{\it (1)}\ Istituto Nazionale per la Fisica della Materia 
and Dipartimento di 
 Fisica, Universit\`a di Cagliari, Italy\\
{\it (2)}\ Department of Physics, University of California,
Davis, U.S.A.  \\
{\it (3)}\ Walter Schottky Institut, Technische Universit\"a{}t M\"u{}nchen,
Germany\\
{\it (4)}\ TECHSO S.p.A., Zona Industriale Est, Cagliari, Italy}
 
\date{\today}
\maketitle
%
\begin{abstract}
Relaxations at nonpolar surfaces
of III-V compounds result from a competition between
dehybridization and charge transfer. First-principles
 calculations for the
 (110) and (10$\overline{1}$0) faces of zincblende and 
wurtzite  AlN, GaN and InN reveal an
anomalous behavior as compared with ordinary III-V 
semiconductors. Additional calculations for GaAs and  ZnO suggest
close analogies with the latter.
We interpret our results in terms of the larger  ionicity (charge
asymmetry) and bonding strength (cohesive energy) in the nitrides 
with respect to other III-V compounds, both essentially due to
the strong valence potential and absence of $p$ core states in the
lighter anion. The same interpretation applies to Zn II-VI compounds. 
\end{abstract}
 
\begin{multicols}{2}
 
\section{introduction}

The III-V nitrides GaN, AlN, and InN are of enormous current 
 interest \cite{diodi}
 in blue optoelectronics and high-power devices technology.
Among the relevant problems in this area, there is the
 high density of
threading dislocations and domain boundary defects
occurring during growth. These boundaries
often  coincide geometrically with the non-polar surfaces of the
material, so that accurate characterizations of these surfaces are of
 interest, and  first-principles calculations in this area
are timely. Although wurtzite
 nitrides are usually grown along the (0001) polar 
direction, other possible growth orientations are being
examined, such as the nonpolar ($10\overline{1}0$) and ($11\overline{2}0$)
surfaces. Also, thin films of zincblende GaN were grown on various 
substrates\cite{crescita} typically along (110) (one of the cleavage
faces of zincblende). 

Earlier works suggest that the nitrides  behave quite  differently than
the ordinary  III-V semiconductors such as GaAs or GaP in several
respects:   the classic gap-cohesive energy relation, \cite{manca}
structural properties,\cite{fmm}  dielectric \cite{bf} and
piezoelectric \cite{bfv} constants. Recent works\cite{jpz,nn} have
pointed out the unusual  surface relaxations of GaN as a further point
of difference. The latter
``anomaly'' would  reflect a stronger ionic character of GaN, making
it similar to the II-VI oxide ZnO, commonly considered as highly
ionic. First-principles calculations\cite{skp,jhh} 
for the (10$\overline{1}$0) surface of ZnO gave smaller rotations
and larger contractions than in GaAs and in other II-VI's
(in view of the 
similar morphology and electronic structure of the (110) and 
(10$\overline{1}$0) surfaces, considerations
about the relaxation mechanism are quite valid for both).
In the present paper we take up this problem for the nitrides,   
 studying the  wurtzite ($10\overline{1}0$) and zincblende
 (110) surfaces of GaN, AlN, and
 InN, and reexamining the properties of the homologous surfaces of ZnO
 and GaAs as reference systems. We discuss the results
 in terms of increased ionicity and increased
cohesive energy of the nitrides as compared to other III-V compounds.
Our interpretation also fits the situation of II-VI compounds,
and is compatible with the presence of the ``anomaly''
only for O and N compounds.
	
The present  first-principles calculations are based on density 
functional theory \cite{dft} in the local density approximation 
(LDA) for the exchange-correlation energy functional,
for which we adopt the Ceperley-Alder \cite{CA} form as parametrized 
by Perdew and Zunger.\cite{PZ}
Ultrasoft pseudopotentials \cite{USPP} have been employed for all the 
elements involved in the calculations.
A plane-wave basis is used  with a cutoff at 25 
Ry. For Ga, In, and Zn, we explicitly include the
semicore  {\it d} electrons in the valence. Slab supercells 
 were used to simulate the surfaces. The results presented here for 
zincblende (wurtzite) surfaces were obtained with 
 symmetric slabs encompassing 8 (9) layers, i.e. 16 (18) atoms, 
whereby all atomic coordinates  were relaxed to obtain forces below
1 mRy/bohr.
A mesh of 10 irreducible special {\bf k}-points  (obtained by
downfolding the bulk mesh)  is used for both	 the
zincblende and the wurtzite surface. 
All calculations are performed at the theoretical lattice constants:
$a$=6.00 bohr, $c/a$=1.613, $u$=0.376 for ZnO;  $a$=10.60 bohr for
GaAs; $a$=5.81 bohr, $c/a$=1.619, $u$=0.380 for AlN;  
$a$=6.04 bohr, $c/a$=1.634, $u$=0.376 for GaN;  
$a$=6.66 bohr, $c/a$=1.627, $u$=0.377 for InN
(see Ref. \onlinecite{bfv} for details on the optimization procedure).

The results for the structural parameters of the zincblende and wurtzite
surfaces are presented in Sec. \ref{sec_fz} and
Sec. \ref{sec_fw} respectively; those for the energetics 
and electronic structure are discussed in  Sec. \ref{sec_enfor}.
In Sec. \ref{sec_inter} we discuss our findings on the
basis of simple chemical concepts.

\section{(110) zincblende surfaces}\label{sec_fz}

\begin{figure}
\epsfxsize=10cm
\centerline{\epsffile{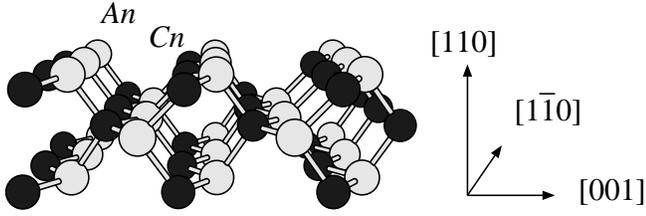}}
\narrowtext
\caption{Side view of (110) ZB surface for III-V semiconductors. 
White spheres are anions ({\it An}), black ones are cations ({\it Cn}).
\label{110_side}}
\end{figure}

\begin{figure}
\epsfxsize=5cm
\centerline{\epsffile{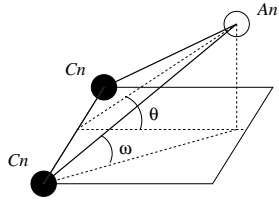}}
\narrowtext
\caption{
Dimer rotation at (110) surface; $\theta$ and $\omega$ are two independent
parameters.
\label{angoli}}
\end{figure}

The relaxations  typical of  the (110) surface of most III-V  
and II-VI compounds have been generally interpreted\cite{qian,bech} as 
driven by a loss of $sp^3$ hybridization towards anion $p$ and cation
$sp^2-$like character. Upon cleavage, charge is transfered from the
cation dangling bond into the anion dangling bond.
The plane containing each anion-cation 
chain running along [1$\overline{1}$0]  rotates with respect 
to the ideal surface (see the sketches in Fig.\ref{110_side}
and Fig. \ref{angoli}).
In each surface dimer,  the cation shifts downwards, so as to 
lay nearly in the plane of its three neighbors, and rehybridizes
to  $sp^2$--like.
The anion shifts upward and is bound to its neighbors by $p$-like 
back-bonds, while it fills up its low-laying $s$-like state.

The relaxations are usually expressed via a combination of
the layer rotation angle  $\theta$, the bond rotation angle $\omega$,
and the bond contraction C$_{\rm B}$ (see Fig.\ref{angoli}). Notice
that
 $\theta$ and $\omega$ are independent parameters, since  the
  dimers can stretch or shorten besides rotating.
Only if  C$_{\rm B}$=0, $\theta$ and $\omega$ are related by 
$\sqrt{3}$\,sin\,$\omega$\,=\,sin\,$\theta$.
In Table \ref{strutt} we list our results for AlN, GaN, InN,
 and GaAs, in comparison with the data of Ref. \onlinecite{jpz} and
(only for 
GaAs) experiments.\cite{bech} The values for GaAs are in very good
agreement with experiment. For nitrides, we do indeed  confirm an
 anomalous behavior: the rotation angles are nearly a half than
for GaAs, and the bond contractions are appreciable, as opposed to
negligible  for GaAs. 
If (see the  discussion in Sec. \ref{sec_inter})
 we interpret small bond rotations and large bond contractions as a
measure of ionicity, we see that the latter grows  along the sequence
InN$\rightarrow$GaN$\rightarrow$AlN, i.e.   inversely with the cation
size.
While qualitatively similar, our results differ somewhat
from those of
Ref. \onlinecite{jpz} for GaN. This discrepancy is probably due
to  the smaller cells and limited relaxations (first layer only) 
in  Ref. \onlinecite{jpz}.  

In Table \ref{spost} the atomic displacements
of first- and second-layer atoms are listed for GaN and GaAs 
($\hat{x}$=[001], $\hat{z}$=[110]). For GaN the largest shifts are 
those of the surface cation, whereas the other atoms shift only slightly
and almost rigidly. In GaAs, the displacements along $\hat{z}$ are
much more relevant for both anion and cation. They move far away
from  each other,  their vertical distance
 [$d$=0.25 $a_0$ (1 -- 0.01 C$_{\rm B}$)
sin $\omega$]   being  0.69 \AA\, in GaAs against 0.23 \AA\, of GaN, 
on account of  a more than double rotation angle, and of course of
the 20 \% 
larger lattice constant.

\section{(10$\overline{1}$0) wurtzite surfaces}
\label{sec_fw}
\begin{figure}
\epsfxsize=9cm
\centerline{\epsffile{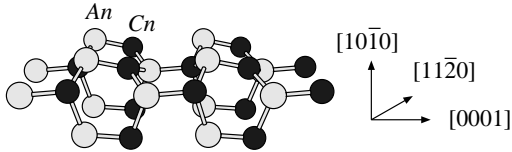}}
\narrowtext
\caption{
Side view of the relaxed (10$\overline{1}$0) surface. White spheres are
anions ({\it An}), black ones cations ({\it Cn}).
\label{wurtz}}
\end{figure}

Wurtzite is the most stable phase of III-V
 nitrides. Its (10$\overline{1}$0) 
surface is sketched in Fig. \ref{wurtz}. At variance with 
zincblende (110), symmetry only allows  dimers rotation
in the plane containing the [10$\overline{1}$0] and [0001] directions,
i.e. ortogonally to the surface plane. Thus there is  only one
rotation angle: $\theta=\omega$. The chemical picture closely
resembles that of the (110). Instead of GaAs, we now consider 
ZnO (10$\overline{1}$0) as reference  system. ZnO  is   
one of the most ionic II-VI  semiconductors, and it allegedly
exhibits
 the same kind of relaxation anomaly \cite{skp}
under examination here for the nitrides;  thus, it represents a
suitable, if  extreme, case for comparison. 

Our results for the relaxations of  wurtzite (10$\overline{1}$0)
(Table \ref{w_rel}) basically confirm the findings for the
zincblende (110) surfaces, with angles and 
bond contractions of the same order of magnitude (angles 
are somewhat smaller and, consistently, contractions are a bit
larger).   For GaN,  previous calculations\cite{nn}
gave comparable, in fact somewhat
 smaller rotation angles. For ZnO our values can be
compared with  theoretical\cite{skp}
and LEED results,\cite{dmpm} and are seen to agree well with the latter.
Among the important features we note 
the close similarity between GaN and
ZnO and  the highly ionic character 
of AlN (see Sec. \ref{sec_inter});
 also, the ``ionicity'' trend InN$\rightarrow$GaN$\rightarrow$AlN
is confirmed. 

In Table \ref{spost_w} we list the atomic displacements
in the first and second surface layer  for GaN and ZnO. For GaN, the
first-layer anions  move upward, the cations downward. The separation
along $\hat{z}$ is 0.36 \AA$\,$ against 0.22 \AA$\,$ 
of Ref. \onlinecite{nn}. As a consequence our $\theta$   is $\sim 40\%$
greater.  Also, we find that even
the second-layer cation moves upward sizeably, whereas in
Ref. \onlinecite{nn} changes in the second layer are
moderate. Finally, for ZnO both surface atoms go down, but their
distance along $\hat{z}$ (0.36 \AA) equals that of GaN.   

\section{Surface  energies and electronic states}
\label{sec_enfor}
In Table \ref{en_sup} we report the surface formation
energies of all compounds studied. $\sigma$ is the surface  energy
per atom of the
 fully relaxed structure, $\Delta \sigma$ the 
energy gained upon relaxation. Our results agree well with   previous
data for GaN (10$\overline{1}$0)\cite{nn} and GaAs
(110).\cite{qian} The formation energy per atom may be  roughly
understood as the energy needed to break a single bond, i.e. 1/4 the
cohesive energy per atom: indeed, at least for the cases  in  Table
\ref{en_sup},  $\sigma$ is close to $E_{\rm coh}/4$.

\begin{figure}
\epsfxsize=9cm
\narrowtext
\centerline{\epsffile{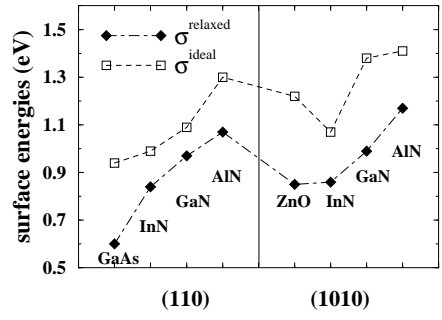}}
\caption{Formation energies $\sigma$ of (110) and (10$\overline{1}$0)
surfaces.}
\label{ensup}
\end{figure}

$\sigma$ is also reported in Fig. \ref{ensup} to make trends easily  
detectable. For the nitrides, the (110) surfaces energies are
$\sim$60\% larger  than in GaAs. This difference is enhanced by
relaxations, that strongly reduce the surface energy of GaAs. For the
(10$\overline{1}$0) surfaces,  energy differences range in an interval
of $\sim$ 0.2 eV. Noticeably, $\sigma$ grows along the same pattern of
ionicity observed previously for bond contractions
(InN$\rightarrow$GaN$\rightarrow$AlN). Finally, the (110) and
(10$\overline{1}$0) surface formation energies   of a given compound
are similar, although the relaxation energy is larger for
the latter. 

\begin{figure}
\epsfxsize=7cm
\centerline{\epsffile{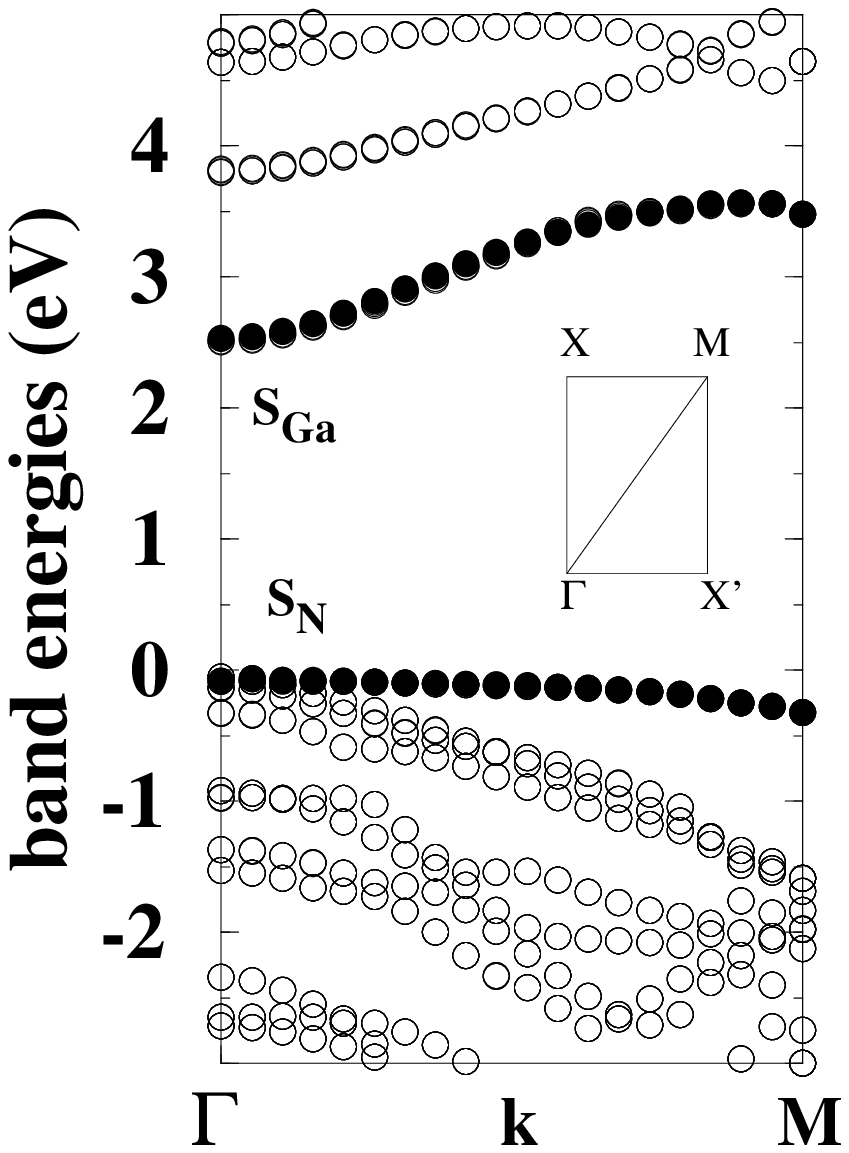}}
\narrowtext
\caption{
Band energies of GaN (10$\overline{1}$0), plotted for {\bf k} 
running along the diagonal of IBZ (shown in the inset). Black circles 
are anionic ($S_N$) and cationic ($S_{Ga}$) surface states.
\label{bande}}
\end{figure}

It is  overall evident that as far as non-polar surfaces are
concerned,  the nitrides are closer to a
highly ionic compound such as ZnO than to GaAs. Similar 
conclusions have been drawn from recent studies on spontaneous 
polarization and piezoelectric constants of bulk nitrides;\cite{bfv}
indeed, all the data suggest that the nitrides are even more extreme
 in their deviation from typical III-V behavior than ZnO 
compared to typical II-VI's.
It is appropriate to check  if such a behavior is also mirrored in
the electronic  properties. Indeed, while no surface states
are present in the gap at the (110) surfaces of
GaAs and GaP\cite{bech} because of dimer rotation,
for ZnO the occupied dangling-bond surface state has been predicted to
lay in the gap.\cite{skp} A previous calculation\cite{nn}
for GaN (10$\overline{1}$0) found the occupied anionic surface state
to lay slightly ($\sim$0.1 eV) below the valence band top. 
We find similar results
(Fig. \ref{bande}), with the anionic surface state touching the
valence top but still remaining within the band edge at $\Gamma$.
 Similar results are found for the other nitrides.
In agreement with the detailed analysis of Ref. \onlinecite{skp} for ZnO,
the  empty surface state corresponding to the remants of the cation
dangling bond is prevailingly $s$--like. The filled  surface state
just above the valence band  corresponds to the 
anion $p$--like back-bonds.
It should also be mentioned  that  recent results by Hirsch {\it et al.}
\cite{hirsch} suggest that the occupied surface states just mentioned
lay indeed completely inside the gap. The difference to ours and other
previous results should be attributed to an improved treatment of the
semicore $d$ states in Ref. \onlinecite{hirsch}.

\section{Discussion}\label{sec_inter}

To describe the  relaxation mechanism, it is useful to consider
separately three items. 

\noindent
{\it First}, on all the (110) surfaces of binary tetrahedrally
coordinated  $A_N B_{8-N}$ compounds,
 a charge transfer occurs from the cation dangling bond into the
anion dangling bond of the as-cleaved surface. This  is a purely
electronic-structure effect, occurring even at zero rotation angle:
the cation dangling bond state is much higher in energy, and it
fully transfers its  electron  occupancy into the   anion state.  
%

\noindent {\it Second}, the surface dimer rehybridizes towards  a
cation $sp^2$--like/anion $p$--like configuration. This entails a
rotation of the dimer (a combined anion-upward,
cation-downwards motion).  This rotation is accompanied by a lowering
of the energy of the occupied anion dangling bond state, and an
increase of the energy of the  cation empty state. This is precisely
the reason why the cation-anion dangling bond occupation transfer is
desirable for this rotation to happen. Since this rehybridization is
qualitatively a kind of reverse of $sp^3$ hybridization, it is
expected to be most favorable when the hybridization energy gain  is
low to begin with. 

\noindent
{\it Third}, the charge within the surface dimer is asymmetric towards
the anion, because of {\it (a)} chemical bond
ionicity, in the spirit of e.g. the  Garcia-Cohen \cite{garcia}
 charge asymmetry, and on top of that {\it (b)} the dangling bond
transfer. 
Therefore the dimer rotation, with the ensuing anion displacement 
out of, and away from the cation plane, costs electrostatic energy.
The energetic cost will be larger, the more asymmetric the charge
distribution is (see e.g. Ref. \onlinecite{monch}).  A key point is now
that, in all materials, there is always a {\it complete} 
cation-anion dangling bond occupation 
transfer: therefore, what matters is the {\it net} anion-cation charge
asymmetry, that is largely equivalent to bulk  
ionicity.\cite{garcia} The larger this is, the less the rotation 
is favored. To be precise,  the dangling bond charge transfer 
will increase the local charge asymmetry (hence hinder rotation) more
strongly in  low-bulk-ionicity compounds:  in the latter,  the bulk
(i.e. pre-cleavage) charge asymmetry is smaller than in more ionic
compounds; thus, when the  full dangling bond occupation is
transferred to the anion,  the net asymmetry increases more than in
strongly ionic compounds. 

We can then  rationalize the energetic balance as follows. 
Rehybridization-plus-rotation is less costly when {\it (a)}
the gain in the reverse process of $sp^3$ hybridization  is low, 
i.e. qualitatively when the cohesive energy of the material is small,
and {\it (b)} when the electrostatic cost of the outward rotation is
low, i.e. when charge asymmetry is small, i.e. ionicity is low.
The predicted trend  is then that materials with small cohesive
energies and ionicities will tend to have large rotations, and viceversa
very ionic and strongly bound solids will tend towards small
rotations. While ``small'' is to be understood in a relative sense,
e.g. for GaAs compared to GaN, or ZnS compared to ZnO, the
nitrides  and ZnO are
 more ionic that all zincblende and wurtzite III-V's and
II-VI's whatever the  ionicity scale. \cite{garcia,monch}
This picture agrees nicely with the calculated quantities for the
nitrides as compared with other III-V's, as well as with those for ZnO
as compared to other II-VI's;\cite{skp}  both ZnO and the nitrides have
both  larger cohesion and ionicity (on any scale),  and
smaller rotation angles  than their companion materials. 
(In the same direction, note that  the  dimer rotation can be
interpreted as a frozen-in zone-center $A_1$ surface  phonon;
\cite{duke} as all other phonons in the nitrides and ZnO, this mode is
stiffer, hence more energetically costly, than in the other III-V's and
II-VI's, respectively.) 

Indeed, large ionicities and cohesive energies, and hence
 small rotations are to be expected
for first-row anions. The basic reason is that
first-row atoms such as N and O have a very deep
potential for the valence $p$ states (and no $p$ core-orthogonality
constraint), whence stronger bonding and larger ionicity than with
other cations. There is indeed a rather abrupt change in rotation
angles (and in other properties too) for first-row anion
both in II-VI's (see CdS vs ZnO 
in Ref.\onlinecite{skp}) and in III-V's; in this sense
selenides are arsenide-like,
sulphides are phosphide-like, and ZnO is GaN-like. 
Interestingly, a similar behavior is observed
in the piezoelectric constants, \cite{bfv}
which increase strongly as the anion decreases in size. For the
nitrides they are large and positive, against the small and negative
values of normal III-V's; for ZnO they are large and positive, against
positive and small in other II-VI's (the II-VI--III-V 
difference is due to a changed balance of the  electronic and ionic
components). 

Note however that one does not expect these trends to hold for any {\it
cation}, in particular small ones. The trend for anions getting
heavier and cations lighter is
towards an effective exchange of roles (e.g. in  boron compounds, this
is reflected in anomalies of structural transitions under pressure),
which blurs the picture somewhat.

\begin{figure}
\epsfxsize=8cm
\centerline{\epsffile{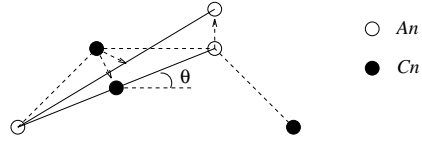}}
\narrowtext
\caption{
Side view of the (110) surface. Dashed line denotes the ideal structure,
full lines two possible atomic rearrangements, one with the anion 
kept fixed in its ideal position, the other with anion shifted
upward.
\label{disegno}}
\end{figure}

Dehybridization and charge 
asymmetry contrast each other also in geometrical terms,
i.e. the larger the layer rotation, the 
smaller the bond contraction. This can be seen by a simple 
geometrical argument. In Fig.\ref{disegno} an idealized picture
of the surface profile is shown. The dashed line refers to the unrelaxed 
surface. If we keep the surface anion fixed and let the cation relax 
onto the plane formed by first- and second-layer anions (thus undergoing
an ideal $sp^3\rightarrow sp^2$ rehybridization), we have
$\theta\simeq$35$^{\circ}$ and a bond contraction of $\sim$5\%. 
If the surface anion relaxes upwards, i.e. towards a more pure 
$p$-like
configuration (which indeed it does),
then  bond contraction tends to be suppressed.
This is the case for GaAs,
where large rotation angles ($\sim 30^{\circ}$) and
small bond contractions ($\sim 1\%$) indicate that dehybridization
dominates (also the case for other III-V's such as GaP).
On the contrary, the small rotations in ZnO and the nitrides are
accompanied by relatively large bond contractions, consistently with
the more critical balance of electrostatic repulsion and
deybridization. 

\begin{figure}
\epsfxsize=8cm
\centerline{\epsffile{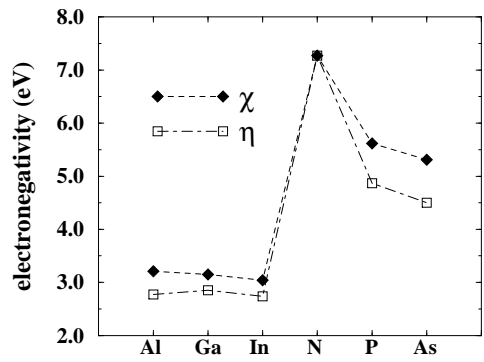}}
\narrowtext
\caption{Electronegativity $\chi$ and hardness $\eta$ for the atomic
components of some III-V semiconductors. N is considerably more
electronegative than  the other anions.}
\label{equal1}
\end{figure}

\begin{figure}
\epsfxsize=8cm
\centerline{\epsffile{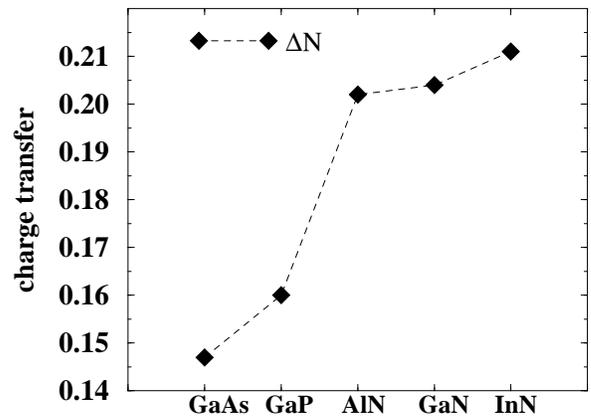}}
\narrowtext
\caption{Charge transfer per dimer given by the electronegativity
equalization  model.} 
\label{equal2}
\end{figure}

It is barely necessary to confirm explicitly that the nitrides are
more ionic than other III-V's.
Charge asymmetry  increases with the electronegativity 
gap between cation and anion, commonly used as measure of
 compound ionicity.
In Fig. \ref{equal1} we report the experimental values of
electronegativity (i.e. one half the sum of atomic ionization potential and
electron affinity) and the hardness (i.e. one half the difference of 
ionization and affinity) for the  atoms under consideration. 
The main feature is that while cations behave quite similarly, this
does not hold for anions,  nitrogen having  much larger values of
 $\chi$ and $\eta$.
How does this influence charge asymmetry in compounds ? 
A semiquantitative estimate is provided by the electronegativity 
equalization principle,\cite{parr-yang} which assumes
the compound energy to be simply the sum of the atomic contributions.
Upon compound formation, one  obtains a charge transfer
\begin{eqnarray}
\Delta N={\chi_B-\chi_A \over 2(\eta_B+\eta_A)}\>,
\label{pino}
\end{eqnarray}
which is depicted in Fig. \ref{equal2}. 
The charge transfer upon   nitride formation
 is much lager, as a consequence of a  greater ionicity.
Use of other ionicity scalesf (Phillips, Pauling, etc.)
will lead to the same qualitative conclusions. For instance
the charge asymmetry coefficients \cite{garcia} $g$\,
 are 0.78, 0.79, and 0.85 for AlN,
GaN, and InN respectively, a factor of $\sim$ 2.5 
larger than the 0.32 of GaN. Indeed, such huge difference 
is  partially mitigated by the large chemical hardness
of N in the denominator  of   Eq. \ref{pino}.

\section{Conclusion}
In summary, non-polar surfaces of III-V nitrides  provide further
evidence that the nitrides are closer to the extreme
ionic limit (embodied e.g. by ZnO) than to normal III-V compounds such
as GaAs, in  agreement with previous results on structural and  
 polarization properties. This strong ionic character  causes
the prevalence of dehybridization in determining surface relaxations
to be less  pronounced than in other III-V's. The same reasoning
applies to ZnO with regard to other II-VI compounds. In the final
analysis, it is   the nature of the nitrogen anion, in particular 
its strong valence potential and the absence of core $p$ states, that
sets the nitrides apart from the other  III-V's, just as the analogous
properties of oxygen cause the major differences of ZnO and other II-V
compounds. 

%
\section*{Acknowledgments} 
The calculations were performed on the IBM SP2  of CRS4 (Centro
Ricerche, Sviluppo,  e Studi Superiori in Sardegna). 
V.F.'s stay
at WSI was supported by the Alexander von Humboldt-Stiftung. 
V.F. thanks Paolo Ruggerone for enlightening
discussions and for reading the manuscript.


\narrowtext
\begin{table}
\refstepcounter{table}
\parbox{\hsize}{TABLE~\ref{strutt} Surface dimer rotation angles 
$\theta$ and $\omega$ (see
Fig. \ref{angoli}), and relative bond contraction C$_{\rm B}$
for zincblende (110) surfaces. $\theta^{\rm a}$, $\omega^{\rm a}$ and 
C$_{\rm B}^{\rm a}$ are from Ref. \onlinecite{jpz}. 
For GaAs experimental values are 
also shown (Ref. \onlinecite{qian}).}
\label{strutt}
\begin{tabular}{ccccc}
  &  AlN & InN & GaN & GaAs \\
\hline
$\theta$   &11.7$^{\circ}$ &14.4$^{\circ}$ &14.3$^{\circ}$ & 30.1$^{\circ}$ \\
$\theta^{\rm a}$ &         &               &2.06$^{\circ}$ & 24.3$^{\circ}$ \\
$\theta^{\rm Expt.}$ &     &               &               & 31.1$^{\circ}$\\
\hline
$\omega$   & 5.8$^{\circ}$ & 7.4$^{\circ}$ & 7.3$^{\circ}$ &16.5$^{\circ}$ \\
$\omega^{\rm a}$ &         &               & 1.0$^{\circ}$ & 13.4$^{\circ}$ \\
$\omega^{\rm Expt}$ &     &               &                & 16.7$^{\circ}$\\
\hline
C$_{\rm B}$    & 5.9\%     &4.3\%          & 4.9\%          & 0.9\%  \\
C$_{\rm B}^{\rm a}$  &     &               & 6.5\%          & 1.3\% \\
C$_{\rm B}^{\rm Expt}$ &  &               &                &  2\% 
\end{tabular}
\end{table}

\begin{table}
\refstepcounter{table}
\parbox{\hsize}{TABLE~\ref{spost}
Displacements from ideal positions (in \AA) of anions ({\it An}) and 
cations ({\it Cn}) in first and second layer of GaN and GaAs (110).
$\hat{x}$=[001] and $\hat{z}$=[110].}
\label{spost}
\begin{tabular}{ccccc}
 & \multicolumn{2}{c}{GaN} & \multicolumn{2}{c}{GaAs} \\
\hline
         &  $\Delta x$  &  $\Delta z$ &  $\Delta x$  &  $\Delta z$ \\
\hline
 $An_1$  &  --0.04      & 0.05      &   0.15   &   0.42  \\
 $Cn_1$  &  0.17        &  --0.18   &   0.37   &   --0.27  \\
 $An_2$  & --0.05       &  0.02     &   0.06   &   0.13   \\
 $Cn_2$ &  --0.03       & 0.07      &   0.08   &   0.23   
\end{tabular}
\end{table}

\begin{table}
\refstepcounter{table}
\parbox{\hsize}{TABLE~\ref{w_rel}
Dimer rotation angle ($\theta$) and relative  bond contraction
(C$_{\rm B}$),  for (10$\overline{1}$0) surfaces. Labels {\it a} and
{\it b} refer to results from Refs. \onlinecite{nn}  and
\onlinecite{skp}, respectively. Experiments are from
Ref. \onlinecite{dmpm}.} 
\label{w_rel}
\begin{tabular}{ccccc}
 & AlN &  InN & GaN & ZnO \\
\hline
$\theta$ & 7.5$^{\circ}$ & 11.0$^{\circ}$ & 11.5$^{\circ}$ & 11.7$^{\circ}$ \\
$\theta^{\rm a}$ &        &                &  7$^{\circ}$   &       \\
$\theta^{\rm b}$ &        &                &                &  7$^{\circ}$ \\
$\theta^{\rm Expt}$ & &    &               &11.47$^{\circ}\pm5^{\circ}$ \\
\hline
C$_{\rm B}$ & 7.5\%   & 5.3\%   & 6.0\%        & 6.0\%      \\
C$_{\rm B}^{\rm a}$ & &         &   6\%         &            \\
C$_{\rm B}^{\rm b}$ & &         &              &  8\%        \\
C$_{\rm B}^{\rm Expt}$ &    &   &              &    --0.9\%  \\
\end{tabular}
\end{table}

\begin{table}
\refstepcounter{table}
\parbox{\hsize}{TABLE~\ref{spost_w}
Atomic displacements of first ($An_1$ and $Cn_1$) and second 
($An_2$ and $Cn_2$) layer from ideal positions (in \AA) for the 
(10$\overline{1}$0) surface of GaN and ZnO. 
($\hat{x}$=[0001] and $\hat{z}$=[10$\overline{1}$0]). {\it An} and 
{\it Cn} indicate 
anion and cation, respectively. Superscript {\it a} indicates 
results from  Ref.\onlinecite{nn}.}
\label{spost_w}
\begin{tabular}{rcccccc}
 & \multicolumn{4}{c}{GaN} & \multicolumn{2}{c}{ZnO} \\
\hline &  $\Delta x$  &  $\Delta x^{\rm a}$ &  $\Delta z$ &$\Delta z^{\rm a}$ 
       &  $\Delta x$  &  $\Delta z$ \\
\hline
 $An_1$  &  0.04    & 0.01    &   0.08     &  0.02    
       &  0.02      &--0.13  \\
 $Cn_1$ & --0.15   & --0.11   &  --0.28    & --0.20   
      & --0.14     &  --0.50    \\
 $An_2$  &   0.04   &  0.05  &  --0.02    &  0.05  
       & --0.02     &  --0.09  \\
 $Cn_2$ &   0.05     & 0.05  &  0.15      &  0.05 
      &   0.03      & --0.09    
\end{tabular}
\end{table}
%
%

\begin{table}
\refstepcounter{table}
\parbox{\hsize}{TABLE~\ref{en_sup}
Surface formation energies ($\sigma$), relaxation energies 
($\Delta \sigma$) 
and cohesive energies per bond (i.e. $E_{\rm coh}$/4, where $E_{\rm coh}$ 
is the cohesive energy per atom). Results are in eV/atom.}
\label{en_sup}
\begin{tabular}{cccccc}
                 & AlN & InN  & GaN  & GaAs  & ZnO\\
\hline
 \multicolumn{6}{c}{(110)} \\
\hline
 $\sigma$        & 1.07& 0.84 & 0.97 & 0.60 &\\     
 $\Delta \sigma$ & 0.23& 0.15 & 0.22 & 0.34 & \\
$E_{\rm coh}/4$  &     &      & 1.09 & 0.81 & \\ 
\hline
 \multicolumn{6}{c}{(10$\overline{1}$0)} \\
\hline
 $\sigma$        & 1.17 & 0.86 & 0.99 & &0.85 \\     
 $\Delta \sigma$ & 0.24 & 0.21 & 0.39 & & 0.37  \\
$E_{\rm coh}/4$  &      &      & 1.09 & & 0.94 
\end{tabular}
\end{table}

%


\end{multicols}
\end{document}